\documentstyle[12pt]{article}

\author{Hao Chen\\
Department of Computing and Information Technology\\
Fudan University, Shanghai 200433\\
P.R. China\\
Liang Ma\\
Institute of Systems Science\\
University of Shanghai for Science and Technology\\
Shanghai 200093, P.R.China\\
and\\
Jianhua Li\\
Department of Electronic Engineering\\
Shanghai Jiaotong University \\
Shanghai 200030, P.R.China}
\title{\bf Vectorial Resilient  $PC(l)$ of Order $k$\\
 Boolean Functions  from AG-Codes}
\date{August, 2006}

\begin{document}

\maketitle
\begin{abstract}
Propagation criteria and resiliency of vectorial Boolean functions
are important for cryptographic purpose (see [1], [2], [3], [4],
[7], [8], [10], [11] and [16]). Kurosawa , Stoh [8] and Carlet [1]
gave a construction of Boolean functions satisfying $PC(l)$ of order
$k$ from binary linear or nonlinear codes. In this paper
algebraic-geometric codes over $GF(2^m)$ are used  to modify the
Carlet and Kurosawa-Satoh's
 construction  for giving  vectorial resilient Boolean functions
satisfying $PC(l)$ of order $k$ criterion. This new  construction is
compared with previously known results.\\


\end{abstract}

{\bf I. Introduction and Preliminaries}\\

In cryptography vectorial Boolean functions are used in many
applications (see [2] and [3]). Propagation criterion of degree $l$
and order $k$ is one of the most general properties of Boolean
functions which has to be satisfied for cryptographic purpose. It
was introduced in Preneel et al [11], which extends the property
strictly avalanche criterion SAC in [16]. For a Boolean function
$f(x)=(x_1,...,x_n)$ of $n$ variables, set
$\frac{Df}{D\alpha}=f(x)+f(x+\alpha)$, $f$ satisfies $PC(l)$ if
$\frac{Df}{D\alpha}$ is a balanced Boolean function for any $\alpha$
with $1\leq wt(\alpha) \leq l$. When the function obtained from $f$
by keeping any $k$ variables fixed satisfies $PC(l)$, we say $f$ has
the property $PC(l)$ of order $k$.  For a vectorial Boolean function
${\bf f}=(f_1(x_1,...,x_n),...,f_m(x_1,...,x_n))$ it is called
$(n,m)-PC(l)$ of order $k$ if any nonzero linear combination of
$f_1,...,f_m$ satisfies $PC(l)$ of order $k$. We say ${\bf f}$
satisfies $SAC(k)$ if it has $PC(1)$ of order $k$ property. A
vectorial Boolean function ${\bf
f}=(f_1(x_1,...,x_n),...,f_m(x_1,...,x_n))$ is called $k$-resilient,
if any nonzero linear combination $\Sigma_i a_if_i$ is a
$k$-resilient. Resiliency of vectorial Boolean functions are
relevant to quantum key distribution and pseudo-random sequence
generators for stream ciphers (see [1], [2], [3], [4] and [17]).\\

We recall the  Maiorana-MacFarland construction of vectorial Boolean
functions. Let $\phi_i : GF(2)^s \longrightarrow GF(2)^r$ be
vectorial Boolean functions for $i=1,...,m$, the class of
Maiorana-MacFarland
 $(r+s, m)$ Boolean functions is the set of the functions $F(x,y)$
 of the form $F(x,y)=(x \cdot \phi_1(y)+h_1(y),...,x \cdot
 \phi_m(y)+h_m(y)): GF(2)^{r+s} \longrightarrow GF(2)^m, (x,y) \in
 GF(2)^r \times GF(2)^s$, where $h_1,...,h_m$ are Boolean functions of
 $s$ variables. It is well known that $F(x,y)$ is at least
 $t$-resilient if $a_1\phi_1(y)+ \cdots +a_m \phi_m(y)$, for any
 nonzero $(a_1,...,a_m) \in GF(2)^m$ and any $y \in GF(2)^s$,  has its Hamming weight at least $t+1$
  (see [1], [2] and [3]).\\

$PC(n)$ Boolean functions of $n$ variables are just the perfect
nonlinear functions introduced by W.Meier and O.Staffebach [10].
They exist only when $n$ is even. Bent functions are the examples of
this kind of functions (see [10] and [16]). People only have few
constructions of $PC(l)$ of order $k$ Boolean functions. In [1] and
[8] $PC(l)$ of order $k$ (vectorial) Boolean functions were
constructed from binary linear or nonlinear codes. For satisfying
the conditions of the construction the minimum distances of the
binary codes and its dual have to be lower bounded. Some lower
bounds on the minimum length (which is the half of the variable
number in the Kurosawa-Satoh construction ) of these binary
linear codes were studied in [9]. \\

From [1] and [8] we know the following results.\\

{\bf Kurosawa-Satoh Theorem ([8]).} {\em Let $C_1$ be a linear
binary code of length $s$ and minimum distance $d_1$ and dual
distance $d_1'$, $C_2$ be a linear binary code of length $t$ with
minimum distance $d_2$ and dual distance $d_2'$. Set
$l=min\{d_1',d_2'\}-1$ and $k=min\{d_1,d_2\}-1$. Then the Boolean
functions of $s+t$ variables
satisfying $PC(l)$ of order $k$ can be explicitly given.}\\

{\bf Corollary 1 ([8] and [9]).} {\em Let $C$ be a linear binary
code with minimum distance at least $k+1$ and dual distance at least
$l+1$. Then Boolean functions of $2n$ variables satisfying $PC(l)$
of order $k$ can be explicitly given.}\\

{\bf  Carlet Theorem ([1]).} {\em For a Boolean function $f(x,y)=x
\cdot \phi(y)+g(y)$ from $GF(2)^{r+s}$ to $GF(2)$, $f$ satisfies
$PC(l)$
of order $k$ if the following two conditions are satified.\\
1) the sum of at least $1$ and at most $l$ coordinates of $\phi$ is
$k$-resilient;\\
2) if $b \in GF(2)^s$ is nonzero and has its weight smaller than or
equal to $l$, at least $k+1$ coordinates of the words
$\phi(y+b)$ and $\phi(y)$ differ.}\\

In this paper the functions $\phi_i$'s in the Mairana-MacFarland
construction are of the form $A_iy+v_i$, where $A_i$ is a fixed $r
\times s$ matrix over $GF(2)$ and $v_i$ is a fixed vector in
$GF(2)^r$, for $i=1,...,m$. \\

Let us now recall some basic facts about AG-codes
(algebraic-geometric codes, see [12], [13] and [14]). Let $X$ be an
absolutely irreducible, projective and smooth curve defined over
$GF(q)$ with genus $g$, $P=\{P_1,...,P_n\}$ be a set of
$GF(q)$-rational points of $X$ and $G$ be a $GF(q)$-rational divisor
satisfying $supp(G)\bigcap P=\emptyset$, $2g-2 < deg(G) <n$. Let
$L(G)=\{f: (f)+G \geq 0 \}$ be the linear space (over $GF(q)$) of
all rational functions with its divisor not smaller than $-G$ and
$\Omega(B)=\{\omega: (\omega) \geq B\}$ be the linear space of all
differentials with their divisors not smaller than $B$. Then the
functional AG-code $C_L(P,G) \subset GF(q)^{n}$ and residual AG-code
$ C_{\Omega}(P,G) \subset GF(q)^{n}$ are defined. $C_L(D,G)$ is a
$[n,k=deg(G)-g+1, d \geq n-deg(G)]$ code over $GF(q)$ and
$C_{\Omega}(P,G)$ is a $[n,k=n-deg(G)+g-1,d \geq deg(G)-2g+2]$ code
over $GF(q)$. We know that the functional code is just the
evaluations of functions in $L(G)$ at the points in $P$ and the
residual code is just the residues of
differentials in $\Omega(G-P)$ at the points in $P$.\\

We also know that $C_L(P,G)$ and $C_{\Omega}(P,G)$ are dual codes.
It is known that for a differential $\eta$ that has poles at
$P_1,...P_n$ with residue 1 (there always exists such a $\eta$, see
[12]) we have $C_{\Omega}(P,G)= C_L(P,P-G+(\eta))$, the function $f$
corresponds to the differential $f\eta$. This means that functional
codes and residue codes are essentially the same.
For many examples of AG codes we refer to [12], [13] and [14].\\

From the theory of algebraic curves over finite fields, there exist
algebraic curves $\{X_t\}$ defined over $GF(q^2)$ with the property
$lim\frac{N(X_t)}{g(X_t)}=q-1$ (Drinfeld-Vladut bound)(see [5] and
[13]), where $N(X_t)$ is the number of $GF(q^2)$ rational points on
the curve $X_t$ and $g(X_t)$ is the genus of the curve $X_t$.
Actually for this family of curves $N(X_t) \geq (q-1)q^t+1$,
$g(X_t)=q^t-2q^{\frac{t}{2}}+1$ for $t$ even and
$g(X_t)=q^t-q^{\frac{t+1}{2}}-q^{\frac{t-1}{2}}+1$ for $t$ odd (see [5]).\\

For a AG-code over $GF(2^m)$ its expansion to some base $B$ of
$GF(2^m)$ over $GF(2)$ will be used in our construction. Let
$\{e_1,..,e_m\}$ be a base of $GF(2^m)$ as a linear space over
$GF(2)$. For a $[n,k,d]$ linear code $C \subseteq GF(2^m)^n$, the
expansion with respect to the base $B$ is a  binary linear code
$B(C) \subseteq GF(2)^{mn}$ consisting of all codewords
$B(x)=(B(x_1),...,B(x_n)), x=(x_1,..,x_n) \in C$. Here $B(x_i)$ is a
length $m$ binary vector $(x_i^1,...,x_i^m)$, where
$x_i=\Sigma_{j=1}^m x_i^j e_j \in GF(2^m)$. It is easy to verify
that the binary linear code $B(C)$ is $[mn,mk, \geq d]$ code. It is
well known that there exists a self-dual base $B$ for any finite
field $GF(2^m)$ of characteristic $2$.
The following result is useful in our construction.\\

{\bf Proposition 1 ([6]).} {\em Let $B$ be a self-dual base of
$GF(2^m)$ over $GF(2)$ and $C$ be a linear code over $GF(2^m)$. Then
the dual
code $B(C)^{\perp}$ is just $B(C^{\perp})$.}\\

A divisor $G$ on the curve $X$ is called effective if the
coefficients of all points in the support $G$ are non-negative. We
say $G_1 \geq G_2$ if $G_1-G_2$ is an effective divisor. This gives
a partial order relation on the set of all divisors. Let
$U_1,...,U_m$ be divisors on the curve $X$, set $max\{U_1,...,U_m\}$
the smallest divisor $U$ such that $U-U_i$ is effective for all
$i=1,...,m$ and $min\{U_1,...,U_m\}$ the biggest  divisor $U'$ such
that $U_i-U'$ is effective for all $i=1,...,m$. For $m$ divisors
$U_1,...,U_m$ and it is clear the intersection $\bigcap_i
L(U_i)=L(min\{U_1,...,U_m\})$, $\bigcap_i \Omega (U_i)=
\Omega(max\{U_1,...,U_m\})$, the linear span of
$L(U_1),....,L(U_m)$ is just $L(max\{U_1,...,U_m\})$.\\

{\bf II. Main Result}\\

The following Theorem 1 and Corollary 2 are the main results of this paper.\\

{\bf Theorem 1.} {\em Let $X$ (resp. $X'$) be a projective,
absolutely irreducible smooth curve of genus $g$ (resp. $g'$)
defined over $GF(2^w)$ (resp. $GF(2^{w'})$), $P$ (resp. $P'$) be a
set of $n$ $GF(2^w)$(resp. $n'$, $GF(2^{w'})$) rational points on
$X$(resp. $X'$), $U_1,...,U_m$(resp. $ U_1',...,U_m'$) be
$GF(2^w)$(resp. $GF(2^{w'}$)-rational  effective divisors on $
X$(resp. $X'$) satisfying $2g-2<deg(max\{U_1,...,U_m\})<n$ and
$supp(max\{U_1,...,U_m\})\bigcap P=\emptyset$ (resp.
$2g'-2<deg(max\{U_1',...,U_m'\})<n'$,
$supp(max\{U_1',...,U_m'\})\bigcap P'=\emptyset$). Suppose
$w(deg(U_i)-g+1)=w'(deg(U_i')-g'+1)$ for $i=1,...,m$. $H$ is another
 $GF(2^{w'})$-rational effective divisor on $X'$ satisfying
$deg(H)+deg(max\{U_1',..,U_m'\}) < n'$ and $w'(deg(H)-g'+1) \geq m$.
It is assumed that $U_1',...,U_m',H$ are disjoint divisors (that is,
their supports are disjoint). Then we have $(wn+w'n',m)$ vectorial
$t$-resilient $PC(l)$ of order $k$ Boolean functions
with $wn+w'n'$ variables, where\\
$$
\begin{array}{ccccccccccccccccc}
l=min\{deg(max\{U_1,...,U_m\})-2g+1, \\
deg(max\{U_1',...,U_m'\})-2g'+1\}\\
k=min\{n-deg(max\{U_1,...,U_m\})-1,\\
n'-deg(max\{U_1',...,U_m'\})-1\}\\
t=n'-deg(max\{U_1',...,U_m',H\})-1.\\
\end{array}
$$
If the curves, the bases of the linear space $L(U_i)$'s and $\Omega(U_i)$'s(resp.
$L(U_i')$'s,  $L(H)$ and $\Omega(U_i')$'s ) are explicitly given,
the $(wn+w'n',m)$ vectorial $t$-resilient $PC(l)$ of order $k$
Boolean functions can be explicitly given.}\\

{\bf Proof.} We consider the linear codes
 $D_1^i=C_L(P,U_i),D_2^i=C_L(P',U_i')$, then
$(D_1^i)^{\perp}=C_{\Omega}(P,U_i),(D_2^i)^{\perp}=C_{\Omega}(P',U_i')$.
Let $B$ and $B'$ be  the self dual bases of $GF(2^w)$ and
$GF(2^{w'}$ over $GF(2)$. We will use the linear binary codes
$C_1^i=B(D_1^i),C_2^i=B'(D_2^i)$. From Proposition 1
$(C_1^i)^{\perp}=B(C_{\Omega}(P,U_i)),(C_2^i)^{\perp}=B'(C_{\Omega}(P',U_i'))$.
The code parameters of $C_1^i$ and $C_2^i$ are $[wn,w(deg(U_i-g+1),
\geq n-deg(U_i)]$ and $[w'n',m'(deg(U_i')-g'+1),\geq n'-deg(U_i')]$.
The code parameters of $(C_1^i)^{\perp}$ and $(C_2^i)^{\perp}$ are
$[wn,w(n-deg(U_i)+g-1),\geq deg(U_i)-2g+2]$ and
$[w'n',w'(n'-deg(U_i')+g'-1),\geq deg(U_i')-2g'+2]$.\\

Let $Q_i$ and $R_i$ be the generator matrices of the binary linear
codes $C_1^i$ and $C_2^i$ respectively, for $i=1,...,m$ . Here we
note that $Q_i$'s (resp $R_i$'s) are $w(deg(U_i)-g+1) \times wn$
matrices (resp. $w'(deg(U_i')-g'+1) \times w'n'$ matrices. Since $w'
(deg(H)-g'+1) \geq m$, we can find  $m$ linear independent vectors
$v_1,...,v_m$ in the binary linear code $B(C_L(H,P'))$. Set
$\phi_i(y)=(R_i)^{\tau}Q_i(y) +v_i, y \in GF(2)^{wn}$ for
$i=1,...,m$, in Maiorana-MacFarland construction we get our
$(wn+w'n',m)$ Boolean function ${\bf f}=(f_1,...,f_m)$. Here
$\phi_i$'s are mappings from $GF(2)^{wn}$ to $GF(2)^{w'n'}$. The
image of $\phi_i$ is the coset $v_i+C_2^i$ for $i=1,...,m$.\\

For any nonzero linear combination $a_1 f_1+...+a_mf_m$, we set
$\phi(y)=\Sigma_i a_i \phi_i(y)+\Sigma_i a_iv_i$. Then it is clear
that $\Sigma_i a_i\phi_i(y)$ is in the binary linear code
$B'(C_L(P',max\{U_1',...,U_m'\}))$
 and $\Sigma_i a_i v_i $ is in the binary linear code
 $B'(C_L(P',H))$. Because $max\{U_1',...,U_m'\}$ and $H$ are disjoint,
 so $\Sigma_i a_i \phi_i(y)+\Sigma_i a_i v_i $ is not zero. On the
 other hand this is a nonzero code word in
 $B'(C_L(P',max\{U_1',...,U_m',H\}))$, its weight is at least
 $n'-deg(max\{U_1',...,U_m',H\})$. Hence ${\bf f}$ is
 $t$-resilient.\\

 From the above argument it is also known that $\phi(y)=\Sigma_i a_i
 \phi_i(y)+\Sigma_i a_i v_i$ is
 in the coset of the binary linear code
 $B'(C_L(P',max\{U_1',...,U_m'\}))$, for any $y \in GF(2)^{wn}$. Thus the sum of arbitrary $j$ (where, $1 \leq j
\leq l$) coordinates $\gamma \cdot \phi(y)$ (here $\gamma \in
GF(2)^{w'n'}, 1 \leq wt(\gamma) \leq l$) of this
 function $\phi(y)$ is a nonzero function, since $l$ is less than the
 Hamming distance of the code
 $B'(C_{\Omega}(P',max\{U_1',...,U_m'\}))
 =(B'(C_L(P',max\{U_1',...,U_m'\})))^{\perp}$. On the other hand
 $\gamma \cdot \phi(y)$ is of the form $u \cdot y +1$ or $u \cdot y$
 (depending on $\gamma \cdot (\Sigma a_iv_i)=1$ or $0$), where $u$
 is a nonzero codeword in $B(C_L(P',max\{U_1',...,U_m'\}))$ with weight at least $k+1$. Thus $\gamma \cdot \phi(y)$ is
 a $k$-resilient function. The 1st condition of the Carlet Theorem is satisfied.\\

 For any $b \in GF(2)^{wn}$, $\phi(y+b)+\phi(y)=\phi(b)$. If $b$ has its weight
  smaller than or equal to $l$, it is
 not  in $B(C_{\Omega}(P,max\{U_1,...,U_m\}))$, thus $Q_ib$ can not
 be zero for all $i=1,...,m$. Thus at least one $(R_i)^{\tau}Q_ib$
 is not zero. From the condition $U_1',...,U_m'$ are disjoint
 effective divisors on $X'$, we know that $\phi(b)=\Sigma_i a_i
 (R_i)^{\tau} Q_i b$ is a nonzero codeword in
 $B(C_L(P',max\{U_1',...,U_m'\}))$. Thus $\phi(b)$ has its weight at
 least $k+1$. The 2nd condition of the Carlet Theorem is satisfied. The
 conclusion is proved.\\

It is well known in the theory of algebraic curves over finite
fields, there are many curves over $GF(2^w)$ (see [12], [13] and
 [14]) with various numbers of rational points and genuses. Thus when
we use Theorem 1 for constructing vectorial $t$-resilient $PC(l)$ of
order $k$ functions, we have very flexible choices of parameters
$l,k, wn+w'n'$. This is quite similar to the role of algebraic
curves in the theory of error-correcting codes. Therefore the
algebraic-geometric method offer us numerous vectorial $t$-resilient
$PC(l)$ of order $k$ functions. Moreover the supports of the
divisors $U_1,...,U_m,U_1',...,U_m',H$ need no to be the $GF(2^w)$
(or $GF(2^{w'})$)
 rational points, it is sufficient the divisors
  are $GF(2^w)$ (or $GF(2^{w'})$-rational. Thus we can easily choose the
  sets of points $P$, $P'$ and the divisors to construct vectorial resilient
  $PC(l)$ of order $k$ Boolean functions. \\

{\bf III. Constructions}\\

In this section  some examples of vectorial $t$-resilient $PC(l)$ of
order $k$ Boolean functions are constructed from Theorem 1 .
Comparing our constructions with the previously known $PC(l)$ of
order $k$ functions in [1] and [8], it seems our constructed
vectorial $t$-resilient $PC(l)$ of order $k$ functions are quite good.\\

We take $X=X'$ the genus $g$ curve which is defined over $GF(2^w)$,
$U_i=U_i'$, $i=1,...,m$, $m$ disjoint effective divisors which are
rational over $GF(2^w)$. In the case $m$ is small and
$deg(U_i)=deg(U_i')=t$ is not 1, we can always choose the supports
of $U_i$'s outside all $GF(2^w)$ rational points on $X$, for
example, we can choose their supports to be $GF(2^{2w})$-rational
points of $X$.  In the following example, $P=P'$ are $n$ $GF(2^w)$
points of $X$. So the only restriction is the upper bound of $n \leq
N(X)$, the number of $GF(2^w)$-rational points of $X$. Because
$U_1,...,U_m$ are disjoint,  $max\{U_1,...,U_m\}=U_1+...+U_m$. Set
$H$ another degree $t'$ $GF(2^w)$-rational effective divisor
satisfying $2g-2<deg(H)<n$ , $w(t'-g+1) \geq m$ , which is supported
on $GF(2^{2w})$-rational points and disjoint to $U_1,...,U_m$ . In
this construction we have $(2wn, m)$ vectorial
$(n-mt-t'-1)$-resilient Boolean functions satisfying
$PC(mt-2g+1)$ of order $n-mt-1$.\\

{\bf Example 1.} We use the genus 0 curve over $GF(4)$ in the
construction. Then $(20,2)$ vectorial $PC(5)$ function is
constructed if we take $m=2,t=2,n=5$.\\

{\bf Example 2.} We use the genus $1$ curve over $GF(4)$ in the
construction, then $n \leq 9$ (see [12] and [14]). We have $(4n, m)$
vectorial $(n-mt-t'-1)$-resilient $PC(mt-1)$ of order $n-mt-1$
Boolean functions, where $2t' \geq m$. Thus $(36,4)$ vectorial
$PC(7)$ Boolean functions are constructed, $(36,3)$ vectorial
$PC(5)$ of order $1$ Boolean functions are constructed, $(24,2)$
vectorial $PC(3)$ of order $1$ Boolean functions  are constructed.\\

When $m=1, t=2$ we have $(n-5)$-resilient $SAC(n-3)$ functions of
$4n$ variables for $n=5,6,7,8,9$.\\

{\bf Example 3.} We use the genus $4$ curve over $GF(4)$ in the
construction, then $n \leq 15$ (see [14]). The $(4n,m)$ vectorial
$(n-mt-t'-1)$-resilient $PC(mt-7)$ of order $n-mt-1$ Boolean
functions  are constructed, where $2(t'-3) \geq m$. Thus we have
$(60,7)$ vectorial $PC(7)$ Boolean functions, $(44,5)$ vectorial
$PC(3)$ Boolean functions, $(48,5)$ vectorial $PC(3)$ of order $1$
Boolean functions,  and $(60,6)$ vectorial $PC(5)$ of order
$2$Boolean functions. \\

When $m=4,t=2$ we have $(4n,4)$ vectorial $(n-14)$-resilient
$SAC(n-9)$ Boolean functions. For example, $(60,4)$ vectorial
$1$-resilient $SAC(6)$ Boolean functions are constructed. \\

{\bf Example 4.} We use the Klein quartic $X$, an algebraic curve
over $GF(8)$ of genus $3$, then $n \leq 24$. From the construction
$(6n,m)$ vectorial $(n-mt-t'-1)$-resilient $PC(mt-5)$ of order
$n-mt-1$ Boolean functions are constructed for $n=7,8,...,24$, where
$3(t'-2) \geq m$. There are at least $19$ degree $2$
$GF(8)$-rational divisors on $X$ (see [14]). Thus we have $(90,7)$
vectorial $PC(9)$ Boolean functions, $(90,6)$ vectorial $PC(7)$ of
order $4$ Boolean functions. When $n=10,...,24$, we have $(6n,3)$
vectorial $(n-10)$-resilient $SAC(n-7)$ Boolean functions.\\

{\bf Corollary 2.} {\em Let $X$ be an algebraic curve over $GF(2^w)$
with genus $g$ and $n$ $GF(2^w)$ rational points and there are at
least $2g$ $GF(2^{2w})$-rational points on $X$. Then we have
$(2wn,g)$ vectorial $(n-\lceil \frac{7g}{2}\rceil-1)$-resilient
$SAC(n-2g-1)$ Boolean functions.}\\

Applying  Theorem 1 to Garcia-Stichtenoth curves [5] over
$GF(2^{2w})$, we have the following result.\\

{\bf Corollary 3.} {\em For positive integers $w \geq 2$ and $h \geq
1$, we have $(4wn,m)$ vectorial Boolean functions satisfying
$PC(mt-2^{2wh+1}+1)$ of order $(n-mt-1)$ for $m$ and $n$ satisfying
$2^{2wh+1}+1 \leq n \leq (2^w-1)2^{2wh}$ and $m \leq n$.}\\

Comparing with the constructions in [1] and [8] we can see our
method based on AG-codes offers more flexibilities for the
parameters $wn+w'n',m,t,k$ and $l$. The main result is more suitable
for constructing {\em vectorial} resilient Boolean functions
satisfying propagation criteria, because there are many
$GF(2^w)$-rational
divisors on the algebraic curves. \\

{\bf IV. Conclusion}\\

In this paper we presented a method based on AG-codes for
constructing $(n,m)$ vectorial
 $t$-resilient Boolean functions satisfying
$PC(l)$ of order $k$ functions . The parameters $n,m,t,k$ and $l$ in
our constructions can be chosen quite flexibly. Many such functions
of less than $100$  variables have been given in our examples.\\

{\bf Acknowledgment.} The work of the first author was supported by
the Distinguished Young Scholar grant 10225106 and grant 90607005 of
NNSF China. The work of the second author was supported by Shanghai
Leading Academic Discipline Project(No.T0502).\\

e-mail: chenhao@fudan.edu.cn\\

\begin{center}
REFERENCES
\end{center}

[1] C.Carlet, On the propagation criterion of degree $l$ and order
$k$, Advances in Cryptology, Eurocrypt'98, LNCS 1403, pages 462-474.\\

[2] C. Carlet  "Boolean Functions for Cryptography and Error
Correcting Codes" (150 pages), chapter of the monography  ``Boolean
methods and models" published by Cambridge University Press (Peter
Hammer et Yves Crama editors).\\

[3] C.Carlet, Vectorial Boolean functions for cryptography, "Boolean
Methods and Models" (Eds Y.Crama and P.Hammer), Cambridge Press.\\

[4] Jung Hee Cheon, Nonlinear vector Boolean functions, Advances in
Cryptology, Crypto 2001, LNCS 2139, pages 458-469.\\

[5] A.Garcia and H.Stichtenoth, On the  asymptotic behaviour of some
towers of function fields over finite fields, J.Number Theory,
61, pages 248-273, 1996.\\

[6] M.Grassl, W.Geiselmann and T.Beth, Quantum Reed-Solomon codes,
in Proc. AAECC 13, LNCS 1719, eds., M. Fossoreier, H.Imai, S.Lin and
A.Poli, Springer-Verlag, pages 231-244, 1996.\\

[7] T.Johansson and E.Pasalic, A construction of resilient functions
with high nonlinearity, IEEE Trans. Inf. Theory, vol. 49(2002), no.
2, pages. 494-501, Feb.2000.\\

[8] K.Kurosawa and T.Satoh, Design of SAC/PC(l) of order $k$ Boolean
functions and three other cryptographic criteria, Advances in
Cryptology, Eurocrypt,97, LNCS 133, pages 434-449.\\

[9] R.Matsumoto, K.Kurosawa, T.Itoh,T.Konno and T.Uyematsu,
Primal-dual distance bounds of linear codes with applications to
cryptography, Cryptology e-print 194/2005, to appear in IEEE Trans. Inf. Theory. \\

[10] W.Meier and O.Staffelbach, Nonlinearity criteria for
cryptographic functions, Advances in Cryptology,
Eurocrypt'89, LNCS 434, pages 549-562.\\

[11] B.Preneel, R.Govaerts and J.Vandevalle, Boolean functions
satisfying high order propagation criteria,
Advances in Cryptology, EuroCrypto'90, LNCS 473, pages 161-173.\\

[12] H.Stichtenoth, Algebraic function fields and codes, Springer,
Berlin, 1993.\\

[13] M.A.Tsfasman and S.G.Vladut, Algebraic-geometric codes, Kluwer,
Dordrecht, 1991\\

[14] G. van der Geer and M. van der Vludgt, Tables of curves with
many points, [Online] Available: http://www.science.uva.nl/~geer/. \\

[15] J.H.van Lint, Introduction to coding theory (3rd Edition),
Springer-Verlag, 1999.\\

[16] A.Webster and S.Tavares, On the design of S-boxes, Advances in Cryptology, Crypto'85, LNCS 218, pages 523-534.\\

[17] X.M.Zhang and Y.Zheng, Cryptographically resilient functions,
vol.43, no.5, pages 1740-1747, Sept.1997.
\end{document}